\documentclass[fleqn,twoside]{article}
\usepackage{espcrc2}

\usepackage{graphicx}

\usepackage{amssymb}

\newlength{\captionspace}
\title{%
\vspace*{-2\normalbaselineskip}
{\tiny DESY 04-158\hspace*{1cm} MS-TP-04-22}\\[\normalbaselineskip]
Unquenched simulations with $N_f=2$ light quark flavours%
}

\author{qq+q Collaboration\\
        Federico Farchioni\address[MS]{Institut f\"ur Theoretische Physik, Universit\"at M\"unster, Wilhelm-Klemm-Str. 9, D-48149 M\"unster, Germany},
        Istv\'an Montvay\address[HH]{Deutsches Elektronen-Synchrotron DESY, Notkestr. 85, D-22603 Hamburg, Germany},
        and Enno E. Scholz\addressmark[HH]%
        \thanks{Talk presented by E. Scholz at Lattice 2004, Fermi National Accelerator Laboratory, June 21-26, 2004.}
}

\begin{document}

\begin{abstract}
The quark mass dependence of meson masses and decay constants in $N_f=2$ QCD is studied at a fixed gauge coupling with different dynamical quark masses. Partially Quenched Wilson Chiral Perturbation Theory (PQW$\chi$PT) is applied to obtain the extrapolation to zero quark mass. It is shown, that in our analysis NNLO-terms play a more important r\^ole than $\mathcal{O}(a)$-terms, which can be neglected. Also we compare our results with recent CP-PACS results.
\vspace{1pc}
\end{abstract}

%
\maketitle
\section{Introduction}
%
To have a reliable extrapolation from lattice-QCD results to physical parameter
values, it is important to simulate (dynamical) quark masses, which are light enough
in that sense that Chiral Perturbation Theory ($\chi$PT) \cite{chpt} can be applied.
If one enters this quark mass region, it is possible to extract the
{\it Gasser-Leutwyler}-constants \cite{gl} from lattice simulations, which then guide
the extrapolation to physical quark masses.
The chiral logarithms predicted by this low energy effective theory may serve as a sign
that one reached sufficiently light quark masses.
\section{Results from PQW$\chi$PT}
Our simulations \cite{qqq_16c32} were done at $\beta=5.1$ with four different sea-quark
(i.e. dynamical) masses.
The lattice size is $L^3\times T=16^3\times32$.
This work is a continuation of \cite{qqq_16to4} were the lattice size was $16^4$ and only
three different sea-quark masses were used.
\par The lattice spacing was found to be $a=0.195(4)\mbox{fm}$, which gives a comfortably
large volume ($L\simeq3\mbox{fm}$).
The dynamical quark masses ($m_{u,d}$) are in the range $\frac14m_s<m_{u,d}<\frac23m_s$,
the pion masses are in $371 \mbox{MeV}\lesssim m_\pi\lesssim 664\mbox{MeV}$.
The extraction of the {\it Gasser-Leutwyler}-constants is done by fitting ratios of masses
and decay constants to the formulas predicted by PQW$\chi$PT \cite{pqchipt}.
We quote our results for the universal low energy scales, assuming $f_0=93\mbox{MeV}$:
\begin{eqnarray*}
\Lambda_3&=&f_0\cdot 8.21(27)=0.76(3)\mbox{GeV},
\\
\Lambda_4&=&f_0\cdot 21.4(1.5)=1.99(14)\mbox{GeV}.
\end{eqnarray*}
These fits also allow to extrapolate the pion decay constant to zero quark mass (for details cf. \cite{qqq_16c32}): 
\[Z_A^{-1}f_0 =121(5)\mbox{MeV}.\]
\par
\begin{figure}[t!]
\begin{center}
\includegraphics*[height=5.5cm,width=5.0cm,angle=-90, bb=50 50 554 604]{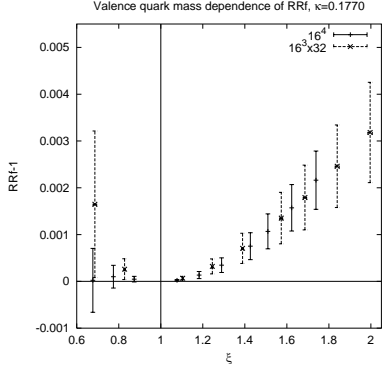}
\end{center}
\vspace{\captionspace}
\caption{Simultaneous plot of $16^4$- and $16^3\times32$-data for $RRf=\frac{f_{VS}^2}{f_{VV}f_{SS}}$ as a function of $\xi=\frac{\chi_V}{\chi_S}$ at $\kappa=0.1770$.}
\label{fig:ratios}
\end{figure}

 Comparing the new $16^3\times32$ data to the previous $16^4$ one, there are deviations
up to 28\% in the measured quantities like $r_0/a$, pion and quark masses
(which shows the largest deviations).
For a more detailed discussion cf. \cite{qqq_16c32}.
It is remarkable that almost all these changes are substantially reduced when comparing
directly the ratios of meson masses and decay constants as functions of ratios of the
(PCAC-)quark masses, fig. \ref{fig:ratios} serves as an example.
\section{$\mathcal{O}(a)$ vs. NNLO}
\begin{figure}[t!]
\begin{center}
\includegraphics*[height=5.5cm,width=5.0cm,angle=-90, bb=50 50 554 604]{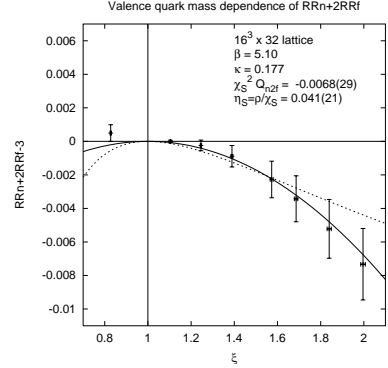}
\end{center}
\vspace{\captionspace}
\caption{Fit of NNLO- (solid) and $\mathcal{O}(a)$-formula (dashed) of $RRn+2RRf-3$.}
\label{fig:dratio_com}
\end{figure}

%
Since we are still working at a fixed gauge coupling $\beta$, from our data we only have indirect information about the discretization effects. That means we did not compare our results at different lattice spacings but fitted the $\mathcal{O}(a)$-effects in the PQW$\chi$PT, which is possible due to the different functional dependence of these terms on the quark mass.
Already in \cite{qqq_16to4} we found the $\mathcal{O}(a)$-contributions in the considered ratios and double ratios, parametrized by
\[
\eta \equiv \frac{\rho}{\chi_S},\;\;\; \rho \equiv \frac{2W_0c_{SW}}{f_0^2}\,a,\;\;\;\chi_S \equiv \frac{2B_0}{f_0^2}\,m_q,
\]
to be small: $\eta\leq0.1$. When trying global fits (i.e. fitting the different ratios and double ratios of the valence mass dependence for all the four sea quark masses together) we still find the best fit at a small $\eta\approx-0.01$ (this would correspond to the effect of a quark mass of $\approx0.5\mbox{MeV}$), but a good fit with $\eta=0$ (no $\mathcal{O}(a)$-terms) is also possible. To confirm this result, we looked at the combination of double ratios
\[
RRn\:+\:2RRf\:-3\:=\left\{\begin{array}{c}\mbox{NNLO}\\\mathcal{O}(a)\end{array}\right. ,
\]
which vanishes in the continuum to NLO order and, therefore, it
could be either described by NNLO- or $\mathcal{O}(a)$-terms.
The two fits are shown in fig. \ref{fig:dratio_com} for our lightest sea quark mass. The $\chi^2$ of the fit is 1.3 for NNLO and 7.2 for $\mathcal{O}(a)$. This special example also suggests that the NNLO is more important in this fitting procedure.
\section{Comparison}
\begin{figure}[t!]
\begin{center}
\includegraphics[height=5.5cm,width=5cm,angle=-90, bb=50 50 554 604]{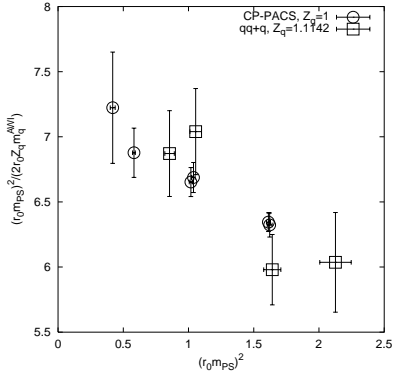}
\end{center}
\vspace{\captionspace}
\caption{Comparison of the ratio $\frac{(r_0m_{PS})^2}{2r_0Z_qm_q}$ vs. $(r_0m_{PS})^2$ between CP-PACS (\cite{cppacs}) and qq+q (\cite{qqq_16c32}).}
\label{fig:comp}
\end{figure}

Other groups are also working (using different fermion actions, finer lattices etc.) on the same problem. For a review cf. \cite{talk_baer}, \cite{talk_ishikawa}.
The CP-PACS collaboration published results at almost the same lattice spacing
($a=0.2\mbox{fm}$) with a set of light sea quark masses at $12^3\times24$ and
$16^3\times24$ \cite{cppacs}.
First we compare the ratio $\frac{(r_0m_{PS})^2}{2r_0Z_qm_q}$ vs. $(r_0m_{PS})^2$
(fig. \ref{fig:comp}).
Because we did not compute any renormalization factor for the quark mass, we fitted our
data to the CP-PACS data, i.e. assuming $Z_q=1$ for CP-PACS.
The best fit gave for the qq+q data: $Z_q\approx1.1142$, which is shown in the plot.
Of course the errors are large, but there is a reasonable agreement between the two data-sets.
\par
\begin{figure}[t!]
\begin{center}
\includegraphics*[height=5.5cm,width=5cm,angle=-90, bb=50 50 554 604]{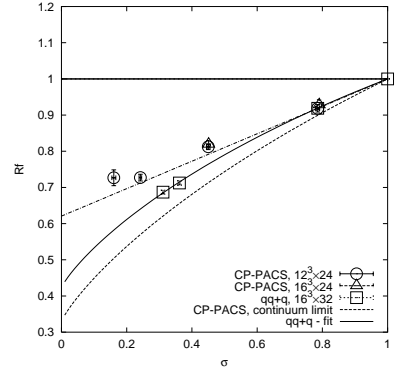}
\end{center}
\vspace{\captionspace}
\caption{$Rf\equiv\frac{f_{SS}}{f_{RR}}$ vs. $\sigma\equiv\frac{\chi_S}{\chi_R}$ for CP-PACS \cite{cppacs} and qq+q \cite{qqq_16c32}. %
Also shown are the (continuum limit) curves and a dashed-pointed straight line connecting the two heaviest qq+q points
}
\label{fig:Rf}
\end{figure}

 Next one may look at the sea-quark mass dependence of $Rf\equiv\frac{f_{SS}}{f_{RR}}$.
By accident it happened that in the two data sets there is a point with almost exactly the
same pion mass $(r_0m_{PS}$): $\kappa=0.14585$ (CP-PACS) and $\kappa=0.1765$ (qq+q).
This is used to set a common scale for comparing the sea-quark mass dependence of the pion decay constant, i.e. looking at the ratio $Rf$. In fig. \ref{fig:Rf} the data points are shown together with the (continuum formula) fit from \cite{qqq_16c32}. The second (dashed) curve shows the fit, if one takes the value of $\Lambda_4=2.44(13)\mbox{GeV}$ from \cite{cppacs} and uses the reference mass parameter $\chi_R=35.8(3.3)$\cite{qqq_16c32}. The qq+q fit gives $\Lambda_4=1.99(14)\mbox{GeV}$, which agrees within errors. The deviation of the CP-PACS data points to the continuum curves should be viewed as (large) $\mathcal{O}(a^2)$-effects. Furthermore the qq+q data-points show the right curvature due to the chiral logarithms, which is easily seen when connecting the two heaviest qq+q points by a straight line (dashed-pointed line in fig. \ref{fig:Rf}).
\section{Discussion \& Outlook}
%
In the previous we showed the higher importance of the NNLO-terms compared to
$\mathcal{O}(a)$-corrections in our data and found a qualitative agreement with recent
CP-PACS results.
\par Recently the phase structure of lattice QCD has been examined by adding a twisted mass
term (\cite{tm_coll} and ref. therein, also \cite{talk_FedericoCarsten}, \cite{tm_frezzotti_talk}).
Besides the Aoki-phase scenario there exists a second scenario which results in a
first order phase transition at $\kappa_{\mbox{\tiny crit}}$.
The possibility that our two lightest quark masses are already inside the metastable
region can not be excluded, as also pointed out in \cite{talk_ishikawa}.
This may be the reason for the observed strong non-linearity of the quark-mass dependence on
$(2\kappa)^{-1}$. We plan to investigate this question in the future.
\par Apart from the study of the phase structure, the twisted mass formulation also
facilitates numerical simulations at light quark masses.
The reason is the lower bound to the eigenvalue spectum provided by the additional mass term.
This suggests to continue these types of simulations in the future with twisted
quark masses.

\begin{thebibliography}{99}
%
\bibitem{chpt} S. Weinberg, Physica {\bf A96} (1979) 327.
%
\bibitem{gl} J. Gasser, H. Leutwyler, Annals Phys. {\bf 158} (1984) 142.
%
\bibitem{qqq_16c32} qq+q-Collaboration, F. Farchioni et al., Eur. Phys. J. {\bf C}
DOI: 10.1140/epjc/s2004-01984-0,  {\tt hep-lat/0403014},
%
\bibitem{qqq_16to4} qq+q-Collaboration, F. Farchioni et al., Eur. Phys. J. {\bf C 31} (2003) 227; {\tt hep-lat/0307002}.
%
\bibitem{pqchipt} G. Rupak, N. Shoresh, Phys. Rev. {\bf D 66} (2002) 054503, {\tt hep-lat/0201019}.
%
\bibitem{talk_baer} O. B\"ar, {\it Chiral Perturbation Theory at Non-Zero a}, plenary talk at Lattice 2004
%
\bibitem{talk_ishikawa} K. Ishikawa, {\it Hadron Spectrum}, plenary talk at Lattice 2004
%
\bibitem{cppacs} CP-PACS Collaboration, Y. Namekawa et al., {\tt hep-lat/0404014}.
%
\bibitem{tm_coll} F. Farchioni et al., {\tt hep-lat/0406039}.
%
\bibitem{talk_FedericoCarsten} F. Farchioni, C. Urbach, parallel talks at Lattice 2004.
%
\bibitem{tm_frezzotti_talk} R. Frezzotti, {\it Twisted-Mass QCD}, plenary talk at Lattice 2004.

\end{thebibliography}
\end{document}